\documentclass[prb,onecolumn,preprintnumbers,amsmath,amssymb,11pt]{revtex4-2}
\usepackage[utf8]{inputenc}
\usepackage{epsf}
\usepackage{graphicx}
\usepackage{latexsym}
\usepackage{amsmath}
\usepackage{amssymb}
\usepackage{amsfonts}
\usepackage{color}

\usepackage{enumitem}

\usepackage{xcolor}
\definecolor{darkblue}{rgb}{0.00, 0.00, 0.55}
\definecolor{darkmagenta}{rgb}{0.50, 0.00, 0.50}
\definecolor{darkcerulean}{rgb}{0.03, 0.27, 0.49}

\usepackage{amsmath}
\usepackage{amssymb}
\usepackage{graphicx,epsf}
\usepackage{hyperref}
\hypersetup{
    colorlinks         = true,
    citecolor          = darkblue,
    linkcolor          = darkblue,
    filecolor          = magenta,
    urlcolor           = darkmagenta,
    bookmarks             = false,
    unicode                  = true,
    bookmarksopen       = true,
    bookmarksopenlevel = 1
}

\begin{document}

\title{\Large{Magnetic force microscopy versus scanning quantum-vortex microscopy:
Probing pinning landscape in granular niobium films}}

\author{A.\,Yu.\,Aladyshkin$^{1-3, *}$, R.\,A.\,Hovhannisyan$^{1}$, S.\,Yu.\,Grebenchuk$^{1}$, S.\,A.\,Larionov$^{1}$, A.\,G.\,Shishkin$^{1}$, O.\,V.\,Skryabina$^{4}$, A.\,V.\,Samokhvalov$^{2,3}$,  A.\,S.\,Mel’nikov$^{1-3}$, D.\,Roditchev$^{5}$, V.\,S.\,Stolyarov$^{1,5}$}

\medskip
\affiliation{$^1$ Moscow Institute of Physics and Technology, 141700 Dolgoprudny, Russia \\
$^2$ Institute for Physics of Microstructures RAS, 603950 Nizhny Novgorod, Russia \\
$^3$ Lobachevsky State University of Nizhny Novgorod, 603022 Nizhny Novgorod, Russia \\
$^4$ Osipyan Institute of Solid State Physics Russian Academy of Sciences, 142432 Chernogolovka, Russia \\
$^5$ ESPCI Paris -- PSL, CNRS, Sorbonne University, France}


\maketitle



\section*{Abstract}

We provide an overview of the methodology and fundamental principles associated with newly developed experimental technique -- scanning quantum-vortex microscopy [Hovhannisyan \emph{et al.}, Commun. Mater., vol. 6, 42 (2025)]. This approach appears promising for experimental studies of vortex pinning phenomena in superconducting films and nanodevices. In particular, we studied the magnetic properties of magnetron-sputtered niobium (Nb) ﬁlms by low-temperature magnetic force microscopy. As the temperature approaches the superconducting critical temperature, the pinning potential caused by structural defects weakens; consequently, the attractive interaction between the magnetic tip of the cantilever and a single-quantum vortex begins to dominate. In this scenario the magnetic probe is capable of trapping a vortex during the scanning process. Because the dragged vortex continues interacting with structural defects, it serves as an efficient nano-probe to explore pinning potentials and visualize grain boundaries in granular Nb films, achieving resolutions ($\sim 30\,$nm) comparable to the superconducting coherence length.

$^*$ Corresponding author, e-mail address: aladyshkin@ipmras.ru

\section*{Introduction}

Defects play an important role in physics of superconductivity \cite{Tinkham,Campbell1972}. The presence of defects in superconducting electronic devices is generally considered undesirable \cite{Sung-18,Golod-18,Grebenchuk-22,Hovhannisyan-24}. On the contrary, structural defects enhance the pinning of Abrikosov vortices, resulting in an increased critical current density for superconducting wires and cables \cite{Golovchanskiy-13,Golovchanskiy-14}.  Disordered superconducting films have high kinetic inductance~\cite{Annunziata_2010,carbillet2020spectroscopic}, making them promising for applications in superconducting quantum devices and sensors~\cite{Soloviev_2021}.

Understanding the physics of the interaction between vortices and structural defects in type-II superconductors constitutes a significant and fundamental challenge \cite{Blatter1994}. Different scenarios of vortex pinning have been intensively studied including pinning on columnar defects~\cite{BerdiyorovPRB2006}, blind holes \cite{BerdiyorovNJP2009}, non-superconducting inclusions \cite{WillaSUST2018}, magnetic particles \cite{Aladyshkin-SUST-review}, among others\cite{KwokRPP2016}. Basic mechanisms of the vortex pinning on structural and magnetic defects apparently include the changes both in energy of supercurrents \cite{BeanPRL1971} and the energy of vortex cores \cite{Blatter1994}. The progress in modern technology enables tuning and controlling the vortex pinning using the sample thickness modulation \cite{DaldiniPRL1974}, substrate engineering \cite{CrisanAPL2001}, surface decoration with magnetic nanoparticles \cite{VanBaelPRB1999}, engineering of peculiar pinning centers \cite{FeighanSUST2017}, and ion irradiation \cite{BugoslavskyN2001}. A detailed knowledge of the pinning network parameters is deeply anticipated in all these cases. However, adequate experimental studies of vortex pinning effects are still challenging. Indeed, pinning potential in real thin-film samples may has different contributions spanning distinct length scales ranging from the superconducting coherence length $\xi$ (tens of nanometers at low temperatures) up to an effective magnetic penetration length $\lambda^{\,}_d=\lambda^2_L/d$ (several micrometers at high temperatures), where $\lambda^{\,}_L$ is the London penetration length and $d$ is the thickness of a sample \cite{Pearl-1964,Pearl-1966}. Thus, an ideal probe should meet the following criteria: (i) nanoscale resolution coupled with a wide field of view; (ii) sensitivity to both bulk and surface properties of superconducting samples; (iii) capability for measurements al low temperatures; (iv) non-destructive regime.

Several microscopy techniques enable the visualization of structural defects at the nanometer scale. Transmission electron microscopy allows for examination at the atomic scale \cite{Devred_2004,golovchanskiy2023magnetization}, however it is a destructive method and examines merely a small fraction of the sample. Scanning probe microscopy techniques such as electron microscopy~\cite{martinez2013microstructures}, tunneling microscopy~\cite{stolyarov2018expansion,berti2023scanning} and atomic-force microscopy~\cite{wu2005studies} also show an excellent spatial resolution and they appear to be non-destructive. However, these approaches reveal defects that protrude at the surface ({\it e.g.} cracks, holes or grain boundaries) and thus provide only limited information about the morphological and electronic properties of defects in the bulk. Specialized low-temperature techniques are employed to investigate specific superconducting properties. Magneto-optical imaging \cite{Veshchunov_2016}, Lorentz microscopy \cite{Tonomura_1992}, magnetic Bitter decoration \cite{Vinnikov_2019}, scanning SQUID microscopy \cite{Embon_2017,halbertal2016nanoscale}, scanning Hall-probe \linebreak microscopy \cite{Moschalkov_2015} and magnetic force microscopy (MFM) \cite{Auslander_2009,Dremov_2019} probe spatial variations of the magnetic field outside the sample. As a result, these methods enable retrieving the distribution of screening (Meissner) and transport currents in the material. Low-temperature scanning laser microscopy \cite{werner2011domain-wall,Werner_2013,Galin_2020} and scanning electron microscopy \cite{Rosticher_2010} make possible experimental investigations of thermal healing processes occurring within a superconductor exposed to localized heating via a focused laser or electron beam \cite{Clem_1980}. These techniques are non-destructive and thus provide useful information about the bulk superconducting properties of cables \cite{Auslander_2009,Embon_2017} and devices \cite{Rosticher_2010,Giovati_2012,Dremov_2019,werner2011domain-wall,Galin_2020}. It is worth to mention that the latter methods have a spatial resolution on the order of micrometers, thereby failing to capture defects at the nanoscale. Therefore, the development of a high-resolution, non-destructive method for probing defect networks in superconductors remains an active area of research.

In this paper, we describe the principles of scanning quantum vortex microscopy (SQVM) and compare the information it provides with that from a more standard MFM approach. As an example, we apply both methods to granular niobium (Nb) films. Such the films are widely used in various applications of superconducting electronics. Several works focused on the vortex pinning on both intrinsic and artificial defects~\cite{dasgupta1978flux,park1992vortex}. The films are known to exhibit a strong vortex pinning on structural defects formed during deposition~\cite{pinto2018dimensional}. While scanning tunneling microscopy and atomic-force microscopy focus primarily on resolving fine details of surface topology, magnetic force microscopy uniquely displays defects located both at the surface and beneath the surface due to a long-range magnetostatic interaction between the magnetic probe and the sample. We use a magnetic probe -- a ferromagnetic-coated tip at a free end of the AFM cantilever -- to either create single-quantum vortices in thin-film samples upon its cooling below the superconducting critical temperature $T^{\,}_c$ or to detect vortices already generated by an external magnetic field. Depending on the relative orientation between the magnetic moment of the probe and the magnetic field above the vortex, an individual vortex may experience interaction with the MFM tip or repulsion away from it. Since the vortex line penetrates entirely through the thickness of the superconducting film, the resultant force exerted to the vortex is a combination of two components describing the interaction to the MFM probe and to various bulk defects. Provided bulk pinning potential is weaker than the vortex-tip interaction, one can consider a scenario, when a single-quantum vortex could potentially be entrapped by the field of the MFM probe and consequently accompany it throughout the scanning process \cite{Hovhannisyan-25, Larionov-25}. During scanning over the studied region of the thin-film sample, the dragged vortex explores the interior part of superconducting film by jumping sequentially from one dominant pinning site to another. These successive jumps can be detected through the modifications of vortex-tip force \cite{Auslander_2009} and then presented in the form of two-dimensional maps, illustrating the spatial variations of the amplitude/frequency/phase of the driven oscillations of the MFM cantilever as a function of the position $(x,y)$ of the magnetic tip. These distributions can be viewed as maps illustrating projections of the hidden defects (like intergrain boundaries in granular films) onto the scanning $x-y$ plane. Since the primary actor in this scenario is the single-quantum vortex dragging behind the MFM probe, this particular imaging mode can appropriately be termed single-quantum vortex microscopy (SQVM), thereby distinguishing it clearly from conventional MFM approach. Remarkably, that the spatial resolution achievable in the SQVM regime ($\sim 35\,$nm) seems to be comparable the superconducting coherence length. It is important to note that both values are much smaller than the expected geometric limitation ($\sim 250\,$nm) typical for standard MFM measurements and conditioned by finite magnetic field penetration length, finite size of the MFM tip's apex, and finite scanning height. We interpret our findings as experimental evidence for vortex core pinning by the grain boundaries.

\section*{Atomic-force and magnetic-force microscopy techniques: elementary theory}

\textsf{\textbf{Atomic-force microscopy.}}  Let's briefly outline the fundamentals of non-contact atomic-force microscopy (AFM) (see, e.g., textbooks\cite{Voig-2015,Voig-2019}).

\begin{figure}[ht!]
\centering
\includegraphics[width=90 mm]{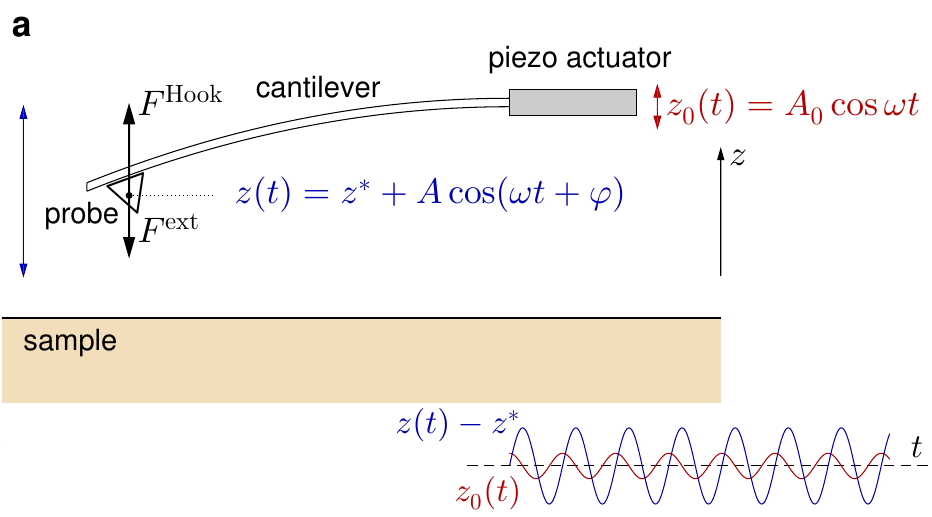}
\includegraphics[width=90 mm]{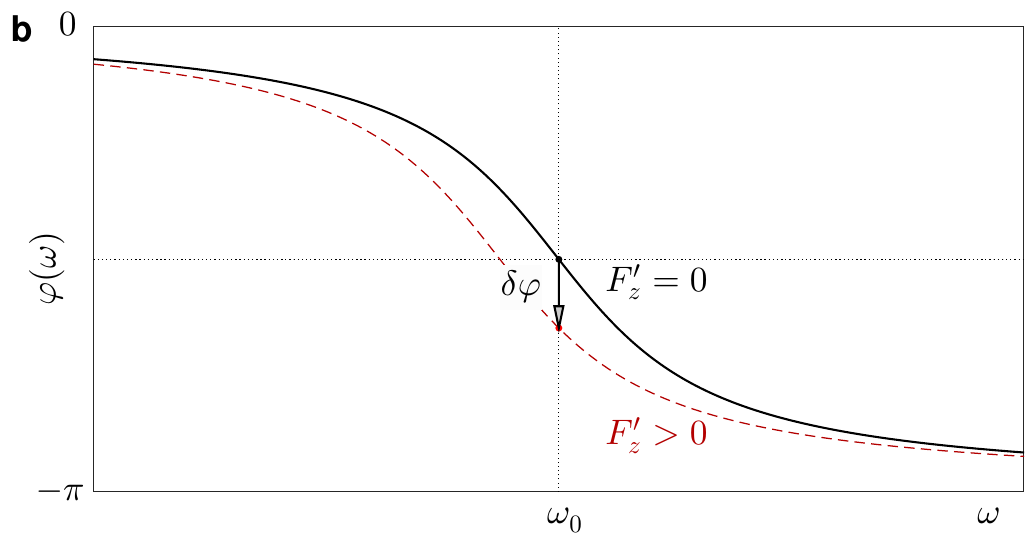}
\caption{\textbf{a} -- Schematic presentation of a cantilever used both in non-contact atomic-force microscopy and magnetic-force microscopy. Inset shows schematically instantaneous displacements of the piezo-actuator (red line) and the probe (blue line); the phase of the probe oscillations lags behind the excitation at about $-\pi/2$.  \textbf{b}~-- Frequency dependence of the phase shift between the excitation and the displacement of the probe.}
\label{Fig01}
\end{figure}

The Newton equation that describes the dynamics of a massive probe (or tip) attached to an elastic beam (or cantilever) can be written as follows
\begin{eqnarray}
\label{Eq:NC-AFM-01}
m^* \ddot{z} = F^{\rm Hook}(z) + F^{\rm ext}(z) - \gamma \dot{z},
\end{eqnarray}
where $m^*$ is the effective mass of the probe, $F^{\rm Hook}(z)=-k\cdot \Delta z$ is the restoring elastic force, obeying the Hook's law, $k$ is the beam stiffness, and $\Delta z = z(t)-z^{\,}_0(t)$ denotes the instantaneous displacement of the probe relative to the fixed end of the cantilever (see figure~\ref{Fig01}a); $-\gamma\,\dot{z}$ represents the frictional force, which is proportional to the velocity of the probe; $F^{\rm ext}(z)$ is the $z$-component of the external force, which accounts for the effects of position-dependent forces between the probe and the sample, as well as position-independent forces such as gravitation and magnetostatic force in a uniform magnetic field.

We assume that the right end of the cantilever (figure~\ref{Fig01}) is harmonically excited by a piezo-actuator
\begin{eqnarray}
\label{Eq:NC-AFM-02}
z^{\,}_0(t)=a^{\,}_0 \cos (2\pi f t) = {\rm Re} \left(a^{\,}_0\,e^{i 2\pi f t}\right),
\end{eqnarray}
where $a^{\,}_0$ and $f$ are the amplitude and linear frequency of the excitations, respectively; ${\rm Re}$ denotes the real part of a complex-valued expression. As a consequence, the free end of the cantilever should undergo harmonic oscillations
\begin{eqnarray}
\label{Eq:NC-AFM-03}
z(t) =  {\rm Re} \left(z^* + a(f)\,e^{i 2\pi f t}\right),
\end{eqnarray}
centered at a specific position $z^*$ [see Eq.~(\ref{Eq:NC-AFM-07})], $a(f)$ is the complex-valued amplitude of the driven oscillations. In the limit of small oscillations, we can expand the position-dependent force into the Taylor-series around the equilibrium point $z^*$
\begin{eqnarray}
\label{Eq:NC-AFM-04}
F^{\rm ext}(z) = F^{\,}_0 + F'_z\cdot (z-z^*),
\end{eqnarray}
where $F^{\,}_0=F^{\rm ext}(z^*)$ is the mean force. Under all these assumptions Eq.~(\ref{Eq:NC-AFM-01}) can be rewritten as follows
\begin{eqnarray}
\label{Eq:NC-AFM-05}
m^* \ddot{z} + \gamma \dot{z}+ (k-F'_z)\,z(t) = k\,z^{\,}_0(t) + F^{\,}_0 - F'_z\,z^*.
\end{eqnarray}
It is easy to see that the introduction of a position-depen-dent force alters the stiffness of the system, which becomes equal to
\begin{eqnarray}
\label{Eq:NC-AFM-05-1}
k^{\,}_{\rm eff} = k - F'_z.
\end{eqnarray}
After substituting Eqs.~(\ref{Eq:NC-AFM-02}) and (\ref{Eq:NC-AFM-03}) into Eq.~(\ref{Eq:NC-AFM-05}) and rearranging the terms, we obtain a simple algebraic complex-valued equation
\begin{eqnarray}
\label{Eq:NC-AFM-06}
\Big\{-m^* (2\pi f)^2 + i \gamma 2\pi f + (k-F'_z)\Big\}\,a(f) e^{i 2\pi f t} +  (k-F'_z)\,z^* = k a^{\,}_0\,e^{i 2\pi f t} + F^{\,}_0 - F'_z\,z^*.
\end{eqnarray}
Since the latter equation should be valid in arbitrary moment, we derive expressions for the equilibrium height
\begin{eqnarray}
\label{Eq:NC-AFM-07}
z^* = \frac{F^{\,}_0}{k},
\end{eqnarray}
and the complex-valued amplitude of the driven oscillations
\begin{eqnarray}
\label{Eq:NC-AFM-08}
a(f) = \frac{f^{2}_0\,a^{\,}_0}{f^{2}_0 - f^2 - f^{2}_0 F'_z/k + i f f^{\,}_0/Q}.
\end{eqnarray}
Here, $f^{\,}_0=(2\pi)^{-1}\sqrt{k/m^*}$ is the resonant frequency of the unloaded cantilever without dissipation,  $Q= 2\pi f^{\,}_0 m^*/\gamma$ is the quality factor of the system, inversely proportional to the dissipation per cycle. For systems exhibiting dissipative probe-sample interactions along with additional energy losses in the sample, the effective $Q$-factor should explicitly incorporate dissipation contributions from both the oscillating cantilever and the sample \cite{Voig-2015}.

\begin{figure}[ht!]
\centering
\includegraphics[width=90 mm]{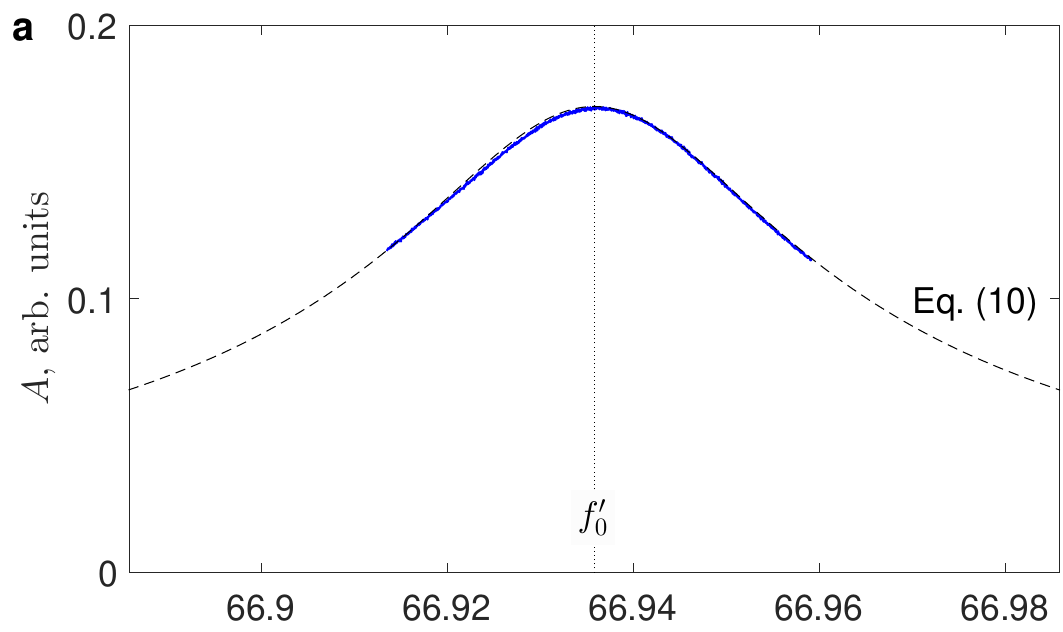}
\includegraphics[width=90 mm]{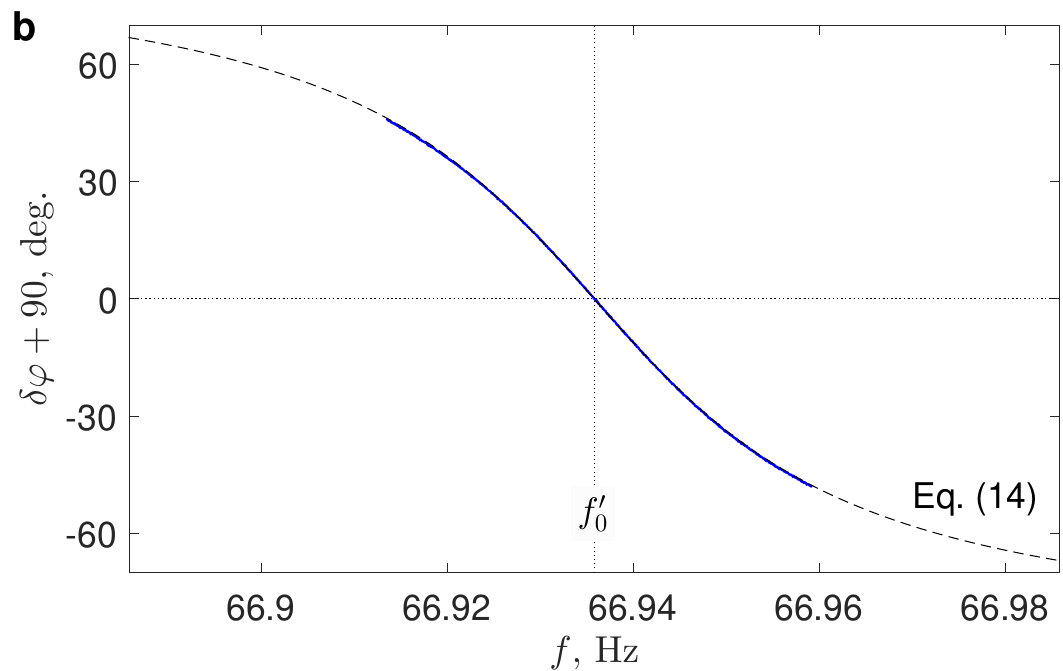}
\caption{Typical frequency dependences of the amplitude of the driven oscillations (panel a) and their phase (panel b) acquired in the MFM measurements at low temperatures. Dashed black lines show the optimal fitting of the experimental data by Eqs. (\ref{Eq:NC-AFM-Ampl}) and (\ref{Eq:NC-AFM-Freq}) with the following parameters: $f^{\,}_0=66.935\,$kHz, $Q=1570$. The phase shift is presented in degrees with respect to an unperturbed baseline value of $-90^{\circ}$.
\label{Fig03-freq-response}}
\end{figure}

Equation (\ref{Eq:NC-AFM-08}) allows us to determine the dependence of the amplitude of the driven oscillations as a function of the frequency of excitation
\begin{eqnarray}
\label{Eq:NC-AFM-Ampl}
A(f) = \left|a(f)\right| =  \frac{f^{2}_0\,a^{\,}_0}{\sqrt{\big(f^{2}_0-f^2 - f^2_0 F'_z/k\big)^2+
\big(f f^{\,}_0/Q\big)^2}}.
\end{eqnarray}
Typical dependence of $A(f)$ for the MFM cantilevers used in this study is shown in figure~\ref{Fig03-freq-response}a. Comparing the experimental frequency response with the model dependence (\ref{Eq:NC-AFM-Ampl}) in the limit of constant force ($F'_z=0$) we can determine the parameters of the cantilever, in particular, the resonant frequency of the loaded cantilever
\begin{eqnarray}
f'_0 \equiv f^{\,}_{0}  \left(1-\frac{1}{4Q^2}\right) \simeq 66.935~{\rm kHz},
\nonumber
\end{eqnarray}
and the quality factor ($Q\simeq 1570$). It is important to note that both energy dissipation and the $Q$-factor strongly depend on the pressure of the exchange gas inside the MFM chamber, leading to potential variations of these parameters between different series of measurements.

One can easily see that the resonant frequency, corresponding to the maximum in the the amplitude-frequency dependence (\ref{Eq:NC-AFM-Ampl}) in the presence of the position-dependent force
\begin{eqnarray}
f^*_0 = \sqrt{f^{2}_0 \left(1-\frac{1}{2Q^2} - \frac{F'_z}{k}\right)} \simeq f^{\,}_{0}  \left(1 - \frac{F'_z}{2k} \right) \quad \mbox{at} \quad Q\gg 1 \quad \mbox{and} \quad |F'_z|/k \ll 1
\label{Eq:NC-AFM-Freq}
\end{eqnarray}
depends on the gradient of the normal component of the external force. The similar expression in the high-$Q$ limit can be straightforwardly derived using the expression (\ref{Eq:NC-AFM-05-1}) for the effective stiffness
\begin{eqnarray}
f^*_0 \simeq \frac{1}{2\pi} \sqrt{\frac{k^{\,}_{\rm eff}}{m^*}} \simeq \frac{1}{2\pi} \sqrt{\frac{k}{m^*} \left(1 - \frac{F'_z}{k}\right)} \simeq f^{\,}_{0}  \left(1- \frac{F'_z}{2k} \right) \quad \mbox{at} \quad Q\gg 1 \quad \mbox{and} \quad |F'_z|/k \ll 1.
\label{Eq:NC-AFM-Freq-eff}
\end{eqnarray}
As a result,  the shift of the resonant frequency in the limit  $Q\gg 1$ is proportional to the first-order derivative of the force $F'_z$ acting between the probe and the sample
\begin{eqnarray}
\delta f \equiv f^*_0 -  f^{\,}_{0} \simeq - f^{\,}_0 \frac{F'_z}{2k}.
\label{Eq:NC-AFM-Freq-2}
\end{eqnarray}

The phase of the driven oscillation in the presence of the position-dependent force is determined as follows (see figure \ref{Fig01}b)
\begin{eqnarray}
\label{Eq:NC-AFM-Phase}
\hspace{-8mm} \varphi(f) = \arg a(f) =
\left\{
\begin{array}{l}
-\arctan \Big(f f^{\,}_0 Q^{-1}/(f^{2}_0-f^2 - f^{2}_0 F'_z/k)\Big) \\
\hspace{36mm} \mbox{at~} f<f^{\,}_0 \sqrt{1- F'_z/k}, \\ [2mm]
-\pi/2  \hspace{27mm} \mbox{at~} f=f^{\,}_0 \sqrt{1- F'_z/k}, \\ [2mm]
-\pi  + \arctan \Big(f f^{\,}_0 Q^{-1}/(f^2 - f^{2}_0 + f^{2}_0 F'_z/k)\Big) \\
\hspace{36mm} \mbox{at~} f>f^{\,}_0 \sqrt{1- F'_z/k}.
\end{array}
\right.
\end{eqnarray}
Typical dependence of the phase on the frequency of excitation is shown in figure~\ref{Fig01}b. It is easy to demonstrate that the phase near the mechanical resonance equals
     \begin{eqnarray}
    \nonumber
    \varphi(f)\Big|^{\,}_{f\simeq f^{\,}_0} = -\pi + \arctan \left(\frac{1}{-Q F'_z/k}\right) \simeq   -\frac{\pi}{2} - \frac{Q F'_z}{k} \quad \mbox{at} \quad \frac{Q |F'_z|}{k}\ll 1.
    \end{eqnarray}
As a result, the measured phase shift becomes proportional to the product of $F'_z$ and the $Q-$factor
    \begin{eqnarray}
    \delta \varphi \equiv \varphi(f) - \left(-\frac{\pi}{2}\right) = - \frac{Q F^{\prime}_{z}}{k}.
    \label{Eq:NC-AFM-Phase}
    \end{eqnarray}

\medskip
\noindent \textsf{\textbf{Magnetic-force microscopy of superconducting samples.}} Magnetic force microscopy (MFM) is a variant of a non-contact AFM technique specifically designed for investigating ferromagnetic, superconducting or hybrid systems \cite{Mironov-2004}. In this case, the dominant interaction is magnetostatic, occurring between the magnetic probe and the stray magnetic field emanating from the sample.

\begin{figure*}[ht!]
\centering
\includegraphics[width=85 mm]{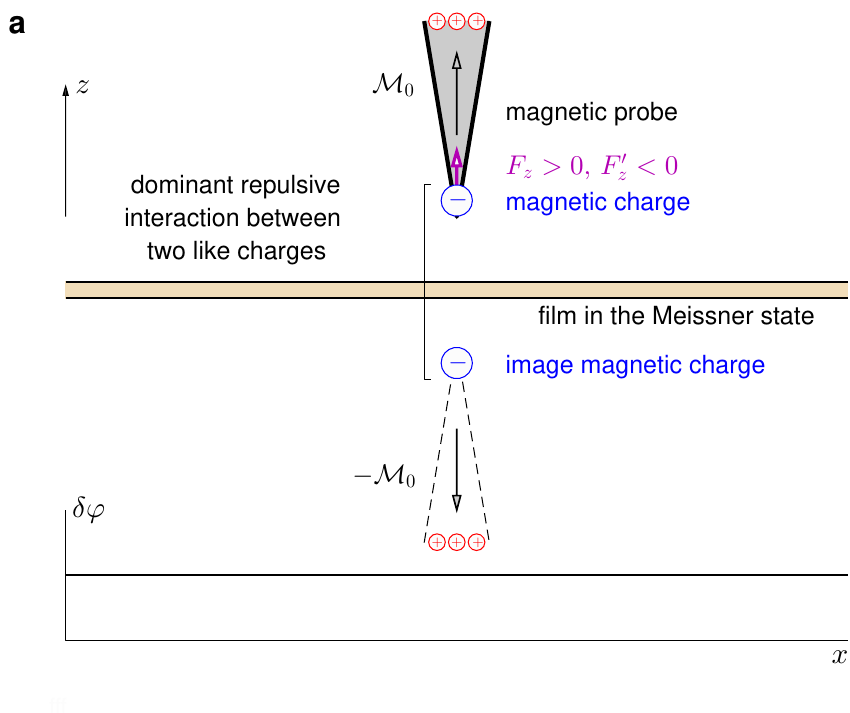}
\includegraphics[width=85 mm]{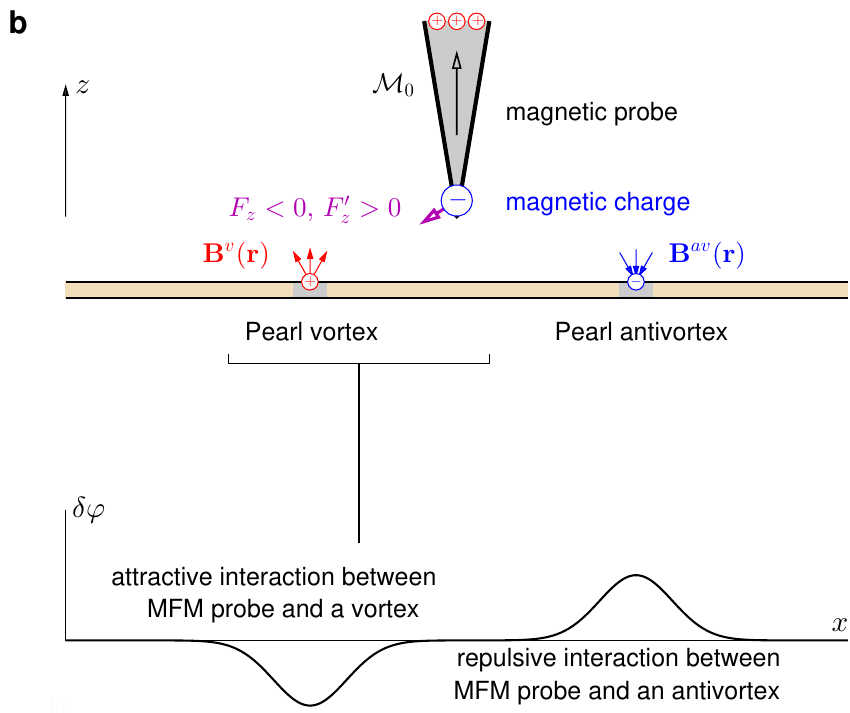}
\caption{\textbf{a} -- Schematic illustration depicting the magnetostatic interaction between an extended magnetic probe and a vortex-free superconducting film due to the ideal Meissner screening. \textbf{b} -- Schematic illustration depicting the magnetostatic interaction between an extended magnetic probe and pinned Pearl vortex/antivortex. Red and blue arrows illustrate the configurations of the magnetic field ${\bf B}({\bf r})$ above a Pearl vortex (on the left) and a Pearl antivortex (on the right). Interestingly, that the ${\bf B}({\bf r})$ distributions resemble the structure of the electric field ${\bf E}({\bf r})$ induced by positive and negative monopoles. This qualitatively explains why the MFM probe attracts to the pinned vortex and repels from the pinned antivortex. The resulting profile of the phase shift $\delta \varphi(x)$ for a superconducting film with pinned vortices should combine both components (uniform and nonuniform), illustrated by thick black lines in the lower parts of panels a and b (see figure~\ref{Fig03-V-and-AV}). }
\label{Fig02-interpretation}
\end{figure*}

We limit our analysis to the effect of an extended MFM probe, whose typical dimensions significantly exceed other relevant length scales such as scanning height $h$ and effective penetration depth $\lambda^{\,}_d$. To facilitate the discussion, we can introduce so called magnetic charges according to conventional definition $\rho^{\,}_m = -{\rm div}\,{\bf M}({\bf r})$, where ${\bf M}({\bf r})$ is the vector of magnetization, corresponding to the probe. For the MFM probe uniformly magnetized along the $z$-axis (figure \ref{Fig02-interpretation}a), the largest magnetic charge, having a negative sign, should be located near the apex of the probe \cite{Hug-1991,Reittu-1992,Giorgio-2019,Hovhannisyan-2021}, while minor positive charges should be located at the distant regions of the probe. The effect of distributed positive charges appears to be significantly weaker compared to that of a localized negative charge (monopole) situated at the minimum distance from the sample.

It is easy to demonstrate \cite{Aladyshkin-99} that that there exists a regime of complete screening in the limit $h\gg \lambda^{\,}_d$. In this case the stray magnetic field induced by the screening currents flowing in thin superconducting film \emph{above} the sample surface is close to the field of the magnetic probe of the opposite polarity, positioned symmetrically with respect to the surface (see dashed line in figure~\ref{Fig02-interpretation}a). It is clear that the magnetostatic interaction between the extended MFM probe and a vortex-free thin superconducting film is equivalent to the repulsive interaction between two closely positioned monopoles of the same signs. Since the force acting on the upper monopole is oriented upward ($F_z>0$) and vanishes as $h$ increases, we conclude that the first-order derivative $F'_z$ should be negative. As a consequence, both the frequency shift $\delta f$ and the phase shift $\delta \varphi$ are expected to remain positive and position-independent because the image monopole adjusts consistently upon relocation of the MFM probe at the given scanning height (see inset in the bottom part of figure~\ref{Fig02-interpretation}a).

If a superconducting film is in the mixed state, the profile of the magnetic field above the sample surface changes drastically due to the presence of vortices. For a Pearl vortex with its magnetic moment aligned parallel to the magnetization of the MFM probe, the local magnetic field ${\bf B}({\bf r})$ above the vortex core points upwards. The ${\bf B}({\bf r})$ patterns mimics the structure of the electric field of a positive charge\cite{Pearl-1964,Pearl-1966}, as depicted schematically in figure~\ref{Fig02-interpretation}b. Thus, the problem of the magnetostatic interaction between the MFM probe and the vortex can be reformulated in terms of electrostatic attraction of two opposite-sign monopoles. Since the $z$-component of the force is negative and vanishes at large distances, we deduce that $F'_z>0$. Consequently, during scanning over the area containing a single well-pinned vortex, the profiles of $\delta f(x,y)$ and $\delta \varphi(x,y)$ are expected to exhibit a distinct minimum precisely above the center of the vortex, tending toward saturation at more distant locations due to the screening effect. For a single-quantum vortex of the opposite polarity commonly referred to as an antivortex we would anticipate a comparable response characterized by a local maximum rather than a minimum (see inset in the bottom part of figure~\ref{Fig02-interpretation}b). Thus, analyzing the spatial distribution of frequency and/or phase shifts allows us to detect the presence of pinned vortices and antivortices.

\section*{Methods}

\noindent \textsf{\textbf{Sample preparation.}} Niobuim films were deposited onto SiO$_2$(270 nm)/Si(001) substrates at room temperature using a rf-magnetron sputtering technique with the following parameters: pre-etching in argon (Ar) plasma during 180\,s (pressure of $2\cdot 10^{-2}$ mbar, mean power of 80\,W at 580\,V),  deposition in Ar plasma (pressure $4\cdot10^{-3}$ mbar, mean power of 200\,W at 238\,V). Typical deposition rate monitored by a quartz microbalance sensor was about 0.2\,nm/s.

\medskip
\noindent \textsf{\textbf{MFM measurement protocol.}} Low-temperature scanning MFM/SQVM measurements were carried out within an AttoCube AttoDry 1000/SU system, equipped with a 9-Tesla superconducting magnet, operating over a temperature range from 4 to 15 K. The setup for the MFM measurements is schematically represented in figure~\ref{Fig02-interpretation}. Both the samples and the MFM cantilever were immersed in a helium buffer gas atmosphere ($\sim 0.5$\,mbar) to ensure rapid thermal equilibration and maintain temperature stability ($\sim 0.3\,$mK). The quality factors of the standard cantilevers may vary from $10^3$ to $5\cdot 10^3$ depending on the pressure of the surrounding He gas and operating temperature. We used regular cantilevers with magnetic Co/Cr-coated probes (MESP, Bruker) with a stiffness constant of 2.8 N/m. Prior to the measurements, the probes were magnetized in the vertical direction in the applied external magnetic field of 2\,kOe at 30\,K (above the critical temperature of the Nb films).

A majority of the experimental data was acquired in a measurement configuration where the frequency-con-trolled feedback loop is deactivated. Operating in this particular mode ensures that both the amplitude and frequency of the mechanical excitations are held fixed throughout the experimentation process. Additionally, the tip height above the sample surface during scanning -- or lift -- is predetermined by the operator and remains unchanged during data acquisition. As a result, we are unable to study topographical features on the sample surface. In this case, we record the map of the spatial variations of the phase shift $\delta\varphi(x,y)$ near the resonant frequency of the loaded cantilever and the amplitude of its oscillations $\delta A(x,y)$.

\begin{figure}[ht!]
\centering
\includegraphics[width=80 mm]{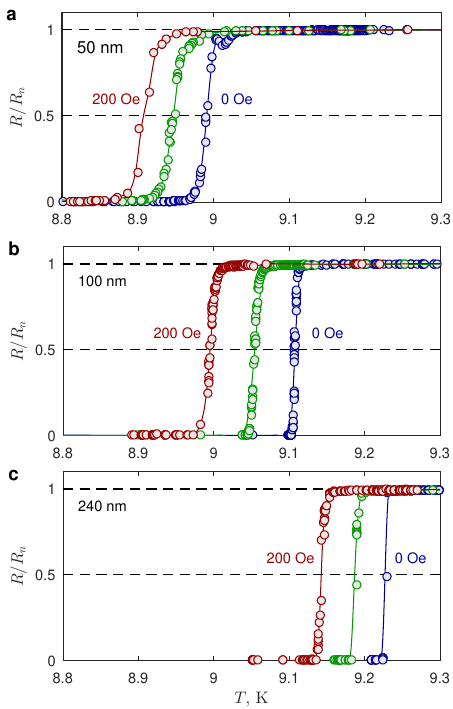}
\caption{\textbf{a, b, c} -- The typical dependences of the normalized dc-resistance $R$ on temperature $T$ for the Nb films of different thicknesses: 50 nm (panel a), 100 nm (panel b), and 240 nm (panel c). Three curves in each panel show the $R(T)$ dependences in the presence of the external magnetic field: zero field (rightmost curves), 100 Oe (middle curves), and 200 Oe (leftmost curve). The critical temperatures, defined at the point corresponding to half of the normal-state resistance are the following: 8.99 K for 50-nm-thick film,  9.11 K for 100-nm-thick film, and 9.23 K for 240-nm-thick film.}
\label{Fig03-resistive-curves}
\end{figure}

In addition, certain portions of the measurements were acquired in the regime with active frequency-controlled feedback via a phase-locked-loop (PLL) technique. In this mode the amplitude of the mechanical excitations is fixed, however the frequency of the excitations is variable. The frequency is automatically adjusted in such a way to maintain zero phase shift (relative to a $-90^{\circ}$ offset) between the excitation and the response of the probe leading to the maximum in the amplitude of the mechanical oscillations. As a result, upon scanning at constant height we can detect spatial variations of the resonant frequency $\delta f(x,y)$ and the amplitude of the oscillations $\delta A(x,y)$.

\section*{Results and discussion}


\noindent \textsf{\textbf{Resistive transition curves.}} Low-temperature transport measurements were performed in an advanced Attocube Attodry 1000/SU system. The sample's resistance $R$ was measured as a function of both temperature and applied magnetic field (oriented perpendicularly to the film surface), using a conventional four-probe configuration. Typical $R-T$ dependences for the Nb films of different thicknesses are shown in figure~\ref{Fig03-resistive-curves}. The observed widths of the resistive transitions are relatively narrow: approximately 0.02 K for 240-nm-thick film and around 0.05 K for 50-nm-thick film. As expected, the external magnetic field shifts all the curves towards lower $T$ values, leaving their shapes almost unchanged. We can estimate the critical temperatures of the samples using the following criterion: $R(T^{\,}_c, H) = 0.5 R^{\,}_n$, where $R^{\,}_n$ is the normal-state resistance of the respective sample. Based on the presented data, we can also estimate the parameter $dT^{\,}_c/dH$, which characterizes the rate of the $T^{\,}_c$ suppression as the field increases from zero up to 200 Oe: $-4.3\cdot 10^{-4}\,$K/Oe for 50-nm and 240-nm thick films and $-5.7\cdot 10^{-4}\,$K/Oe for 100-nm thick film. A 25\% discrepancy between these two values could potentially arise due to uncontrollable variations in the fabrication processes specific to these particular samples. Using linear extrapolation formula, we can estimate the limiting values for the upper critical field and the superconducting coherence length at zero temperature for our samples:
\begin{eqnarray}
\label{zero-temp-estimates}
H^{(0)}_{c2} \simeq \frac{T^{\,}_c}{\left(dT^{\,}_c/dH\right)_{T\to T^{\,}_c}} \lesssim 22.5\,{\rm kOe}, \quad
\xi(0) \simeq \sqrt{\frac{\Phi^{\,}_0}{2\pi H^{(0)}_{c2}}} \gtrsim 12\,{\rm nm}.
\end{eqnarray}
Taking $\xi(0)\simeq 12\,$nm and considering it as the coherence length in the dirty-limit, we can estimate mean-free path $\ell \simeq \xi^2(0)/\xi^{*}(0)\simeq 3.6\,$nm, where $\xi^{*}(0)\simeq 39\,$nm is the zero-temperature coherence length in the clean limit \cite{Grigoriev}. Correspondingly, the London penetration depth in the dirty-limit approximation at zero temperature can be estimated as $\lambda^{\,}_L(0) = \lambda^{*}_L \sqrt{\xi^{*}(0)/\ell}\simeq 130\,$nm, where $\lambda^{*}_L(0)\simeq 38\,$nm is the zero-temperature London penetration length in the clean limit \cite{Grigoriev}.

We realize that the values $\xi(0)$ and $\lambda^{\,}_L(0)$ can be considered as order-of-magnitude estimates. Nevertheless, they allow us to assume that the temperature-dependent coherence length and London length
\begin{eqnarray}
\label{xi}
\xi = \frac{\xi(0)}{\sqrt{1-T/T^{\,}_c}} \quad \mbox{and} \quad \lambda^{\,}_L = \frac{\lambda^{\,}_L(0)}{\sqrt{1-T/T^{\,}_c}}
\end{eqnarray}
at $T\sim 7.5\,$K (the threshold value for a switching between the MFM and SQVM regimes as as demonstrated later) are close to 30\,nm and 300\,nm, respectively. Since the London penetration depth in this temperature range is larger than the thickness of the Nb films (240 nm maximum), we would use the term 'Pearl vortex' to refer to a single-quantum vortex. \cite{Pearl-1964,Pearl-1966}

\medskip
\noindent \textsf{\textbf{240-nm thick Nb film.}} We start with the examining the magnetic response of rather thick superconducting film characterized by an average thickness of 240 nm and the critical temperature  $T^{\,}_c=9.23\,$K in zero magnetic field (figure~\ref{Fig03-resistive-curves}c).  Throughout the paper, the horizontal axis on all two-dimensional images corresponds to the fast-scanning direction.

\begin{figure}[ht!]
\centering
\includegraphics[width=85 mm]{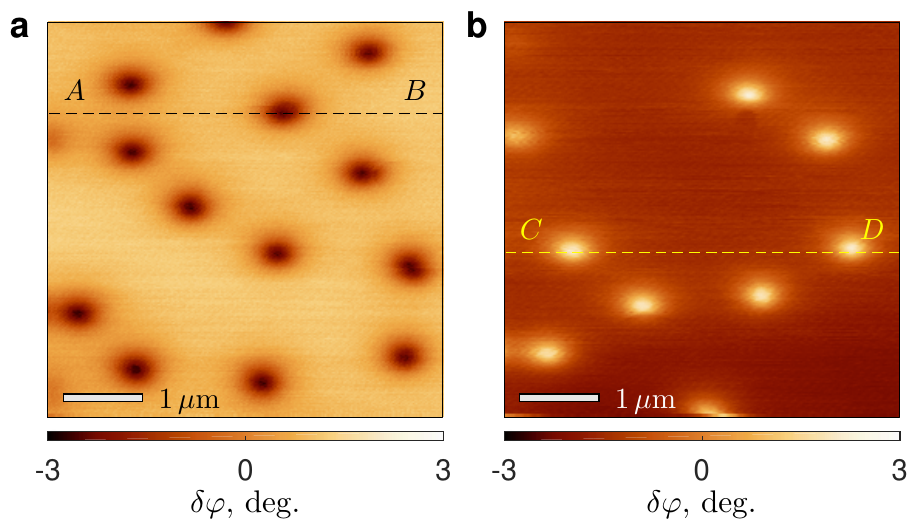}
\includegraphics[width=85 mm]{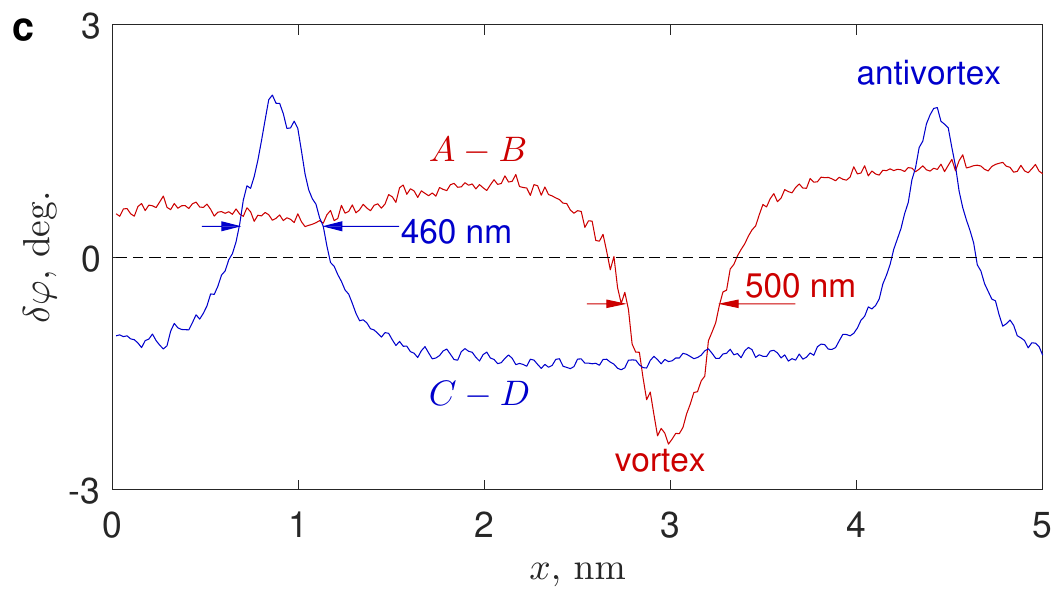}
\caption{\textbf{a, b} -- Spatial distributions of the phase shift $\delta\varphi(x,y)$ in 240-nm-thick Nb film at low temperatures showing ensembles of the vortices and antivortices trapped in 240-nm thick Nb film after cooling down in the presence of the external magnetic field of $+10\,$Oe and $-10\,$Oe, respectively (image size $5\times 5\,\mu$m$^2$, $T=4.06\,$K, scanning height 80 nm, regime of the constant frequency 66.941 kHz). \textbf{c} --  Profiles of the phase shifts $\delta\varphi(x,y)$ along the $A-B$ line (panel a) and the $C-D$ line (panel b).}
\label{Fig03-V-and-AV}
\end{figure}

Figure \ref{Fig03-V-and-AV}a,b shows two phase-shift maps of vortices created in the superconducting film by an external magnetic field at low temperatures ($T/T^{\,}_c\simeq 0.44$). The reason why the vortices do not follow the moving MFM probe relates to the fact under the specified experimental conditions -- such as temperature, scanning height, and tip magnetization -- the pinning force acting on the vortex from the various defects in the sample exceeds the attractive force originating from the MFM probe. For the sake of brevity we will refer to such maps displaying pinned vortex/vortices as the MFM maps. Important to note that in this particular experiments the Nb film was cooled down in the presence of the applied magnetic field (so-called field-cooled process). The magnitude of the field (10\,Oe) is sufficient to create several magnetic flux quanta in the area of interest. Below $T^{\,}_c$, the trapped magnetic flux becomes quantized.  Depending on the orientation of the external magnetic field (up/down), single-quantum vortices appear as dark and bright spots of the same sizes/intensities (panels a and b, respectively). Spatial patterns in the $\delta\varphi(x,y)$ maps are consistent with qualitative arguments (figure~\ref{Fig02-interpretation}b). As anticipated, vortices display a tendency to arrange themselves into a hexagonal lattice; however the resulting positions of individual vortices are unpredictable due to various structural imperfections. Note also, that a single-quantum vortex can be formed in a desirable location by cooling the superconducting film below its critical temperature in the presence of the stray magnetic field generated by an MFM probe positioned in a close vicinity of the sample surface.

\begin{figure*}[ht!]
\centering
\includegraphics[width=85 mm]{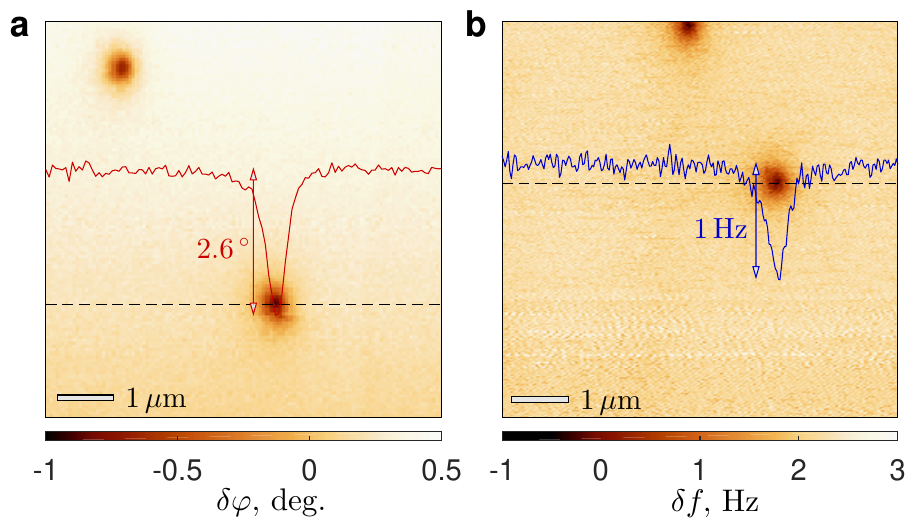}
\includegraphics[width=85 mm]{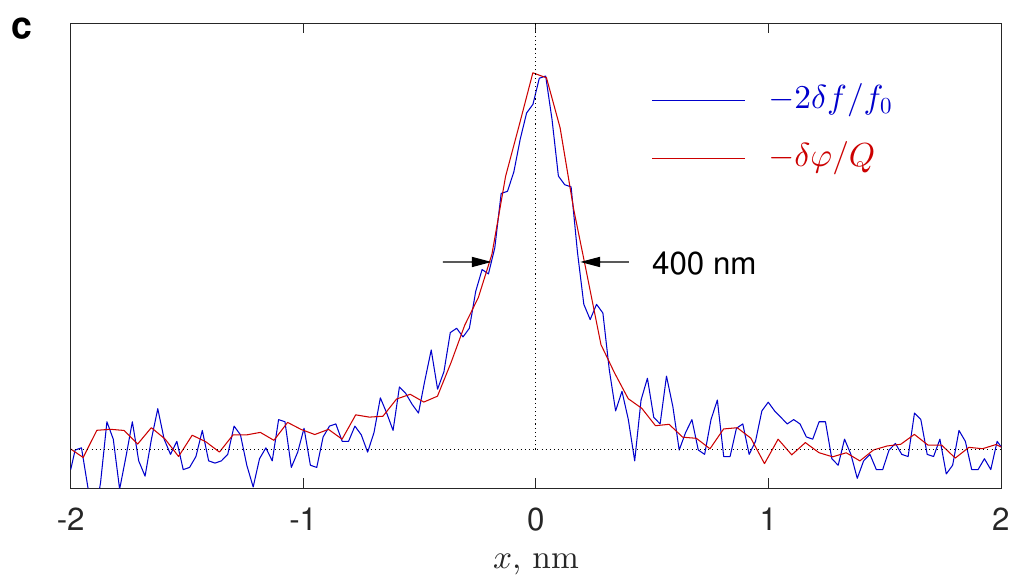}
\caption{\textbf{a} -- Spatial distribution of the phase shift $\delta \varphi(x,y)$  for 240-nm thick Nb film at moderate temperatures recorded with disabled frequency-control feedback (image size $7\times 7\,\mu$m$^2$, $T=7.85\,$K, scanning height 100\,nm, measurement frequency 66.936\,kHz). \textbf{b} -- Spatial distribution of the shift of the resonant frequency $\delta f(x,y)$ for the same film recorded with active frequency-control feedback (image size $7\times 7$ nm$^2$, $T=7.17\,$K, scanning height 100\,nm, mean resonant frequency 66.941\,kHz). \textbf{c} -- Comparison of the normalized profiles $-\delta\varphi(x)/Q$ (along the dashed line in panel a) and $-2\delta f(x)/f_0$ (along the dashed line in panel b) and allows us to estimate the quality factor ($Q\simeq 1500$). }
\label{Fig-240-df-dphi}
\end{figure*}

Sectional views of the phase-shift patterns along the dashed lines in Fig.~\ref{Fig03-V-and-AV}a,b are illustrated as red and blue curves respectively in Fig.~\ref{Fig03-V-and-AV}c. It demonstrates that the visible 'vortex size' (full width at half maximum) is about 460 nm at 80\,nm lift. Similar spot sizes (about 400\,nm at 100\,nm lift) can be derived by analyzing the vortex patterns in Fig.~\ref{Fig-240-df-dphi}c. Such large values of the spot sizes are attributed by the effect that magnetic fields from the Pearl vortices spread as they approach the superconductor-vacuum surface from within the superconductor \cite{Pearl-1964,Pearl-1966,Kirtley}. This means that the spatial extent of the magnetic fields in the lateral direction above the surface is larger than in bulk superconductors. The influence of the stray fields should be taken into account when estimating penetration depths via magnetic imaging techniques.

Figure~\ref{Fig-240-df-dphi}a shows the map of the spatial variations of the phase shift $\delta \varphi(x,y)$ acquired in the regime with deactivated frequency-controlled feedback at constant lift. Figure~\ref{Fig-240-df-dphi}b displays the map illustrating the spatial variations of the resonant frequency $\delta f(x,y)$ acquired in the regime with active frequency-controlled feedback at constant height. Two dark spots  in these maps can be viewed as fingerprints of two individual Abrikosov vortices presumably induced by the stray magnetic field of the Earth \cite{Comment}. Important to note that at $T\simeq 7.8\,$K (\emph{i.\,e.} $T/T^{\,}_c\simeq 0.84$) the vortices can be still pinned provided that the MFM probe is positioned at rather large distance from the sample's surface (100\,nm).

 \begin{figure*}[ht!]
\centering
\includegraphics[width=150 mm]{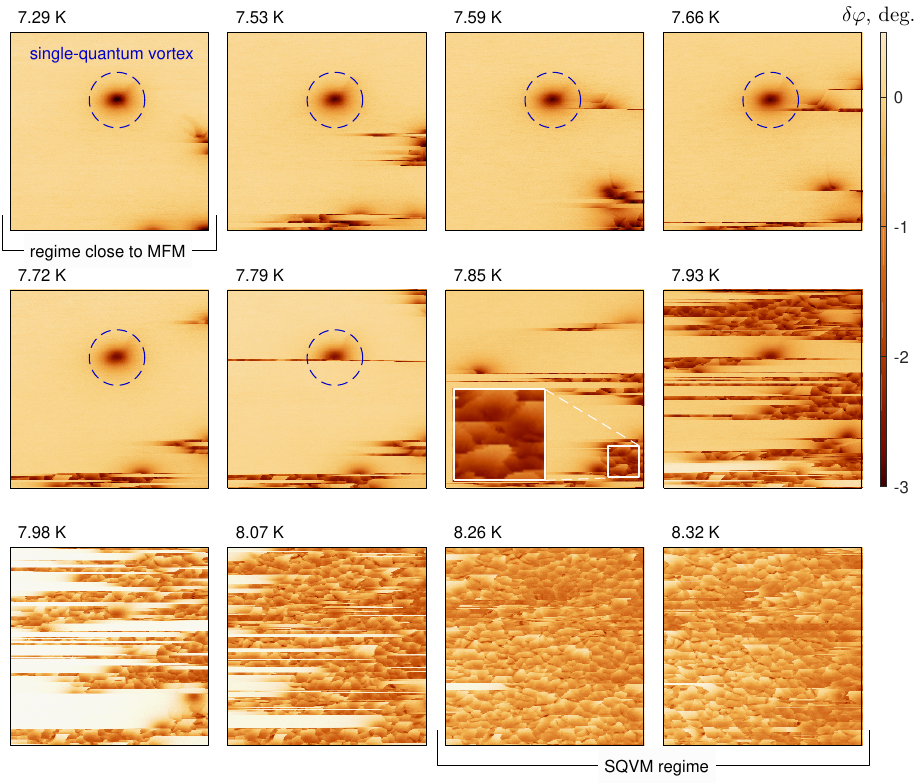}
\caption{Evolution of the phase-shift patterns $\delta\varphi(x,y)$ for the same area of 240-nm-thick Nb film with changing temperature (image size $5\times 5\,\mu$m$^2$, scanning height 40 nm, regime of the constant frequency 66.936 kHz). In fact, all data were recorded sequentially starting from higher temperatures toward lower values.}
\label{Fig04-evolution}
\end{figure*}

\begin{figure*}[h!]
\centering
\includegraphics[width=170 mm]{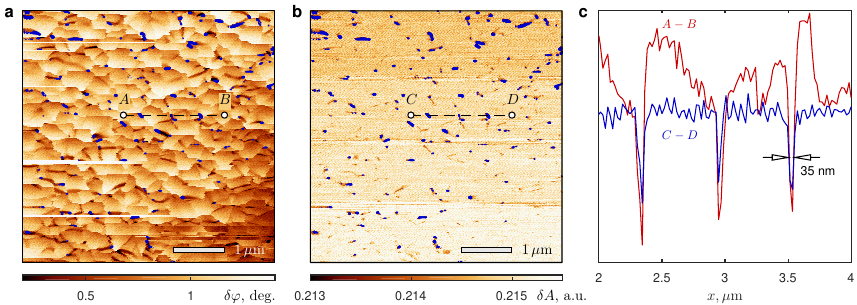}
\caption{\textbf{a, b} -- Spatial distributions of the phase shift $\delta\varphi(x,y)$ (panel a) and the amplitude of the driven oscillations $\delta A(x,y)$ (panel b) for the same area of 240-nm-thick Nb film, acquired at high temperatures (image size $5\times 5\,\mu$m$^2$, $T=8.563\,$K, scanning height 80 nm, regime of the constant frequency 66.936 kHz). For a more effective comparison between these maps, we have highlighted the local minima in the amplitude $\delta A(x,y)$ in each panel (blue lines). \textbf{c} -- Profiles of $\delta\varphi(x)$ taken along the $A-B$ line (red) and $\delta A(x)$ taken along the $C-D$ line (blue). The lines depicted here have been intentionally drawn out of scale and shifted vertically to enhance visual clarity. Small-scale variations in $\delta A(x)$ and $\delta \varphi(x)$ appear to represent noise rather than the prominent minima linked to effective pinning centers.}
\label{Fig04-comparison}
\end{figure*}

According to Eqs.~(\ref{Eq:NC-AFM-Freq-2}) and (\ref{Eq:NC-AFM-Phase}), both $\delta f/f^{\,}_0$ and $\delta \varphi$, measured in two different modes, are proportional to the same geometrical factor $F'_z({\bf r}^{\,}_0)$, where ${\bf r}^{\,}_0$ is the position of the apex of the MFM probe. This makes possible to estimate independently the $Q$-factor of the cantilever. Indeed, the profiles of $\delta f(x,y)$ and $\delta \varphi(x,y)$ along the lines running via the centers of the vortices (figure \ref{Fig-240-df-dphi}c) after normalization have the same shape, therefore
    \begin{eqnarray}
    Q \simeq \frac{1}{2}\cdot\frac{\max |\delta \varphi|}{\max \left|\delta f/f^{\,}_0\right|} \simeq  \frac{1}{2}\cdot \frac{2.6\cdot (\pi/180)}{1/(66.9\cdot 10^3)} \simeq 1.5\cdot 10^3.
    \end{eqnarray}
The factor $\pi/180$ serves to convert the phase shift expressed in degrees into the equivalent value in radians. It is easy to see that the estimated $Q$-value is close to that found previously by the fitting of the frequency responses (figure~\ref{Fig03-freq-response}).

\begin{figure}[ht!]
\centering
\includegraphics[width=90 mm]{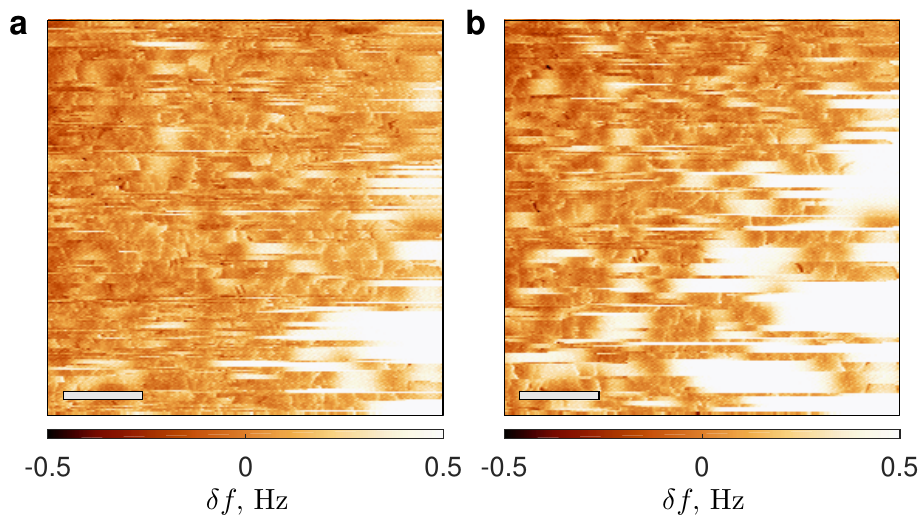}
\caption{\textbf{a, b} -- Spatial distributions of the frequency shift $\delta f(x,y)$ in 240-nm-thick Nb film measured in the regime with active frequency-controlled feedback (image size $5\times 5\,\mu$m$^2$, $T\simeq 8.05\,$K (panel a) and $7.92\,$K (panel b), scanning height 100 nm, mean resonant frequency 66.916 kHz). Scale bars correspond to $1\,\mu$m. Both panels correspond to an intermediate regime, however the fish-scale pattern is clearly visible.}
\label{Fig13-df}
\end{figure}

Series of the $\delta\varphi(x,y)$ maps in figure \ref{Fig04-evolution} demonstrate a crossover from vortex pinning on internal defects to vortex depinning as temperature increases. At the lowest presented temperature (7.29\,K) we clearly see a single vortex whose shape and location are not influenced by the MFM probe moving in the forward and backward direction during scanning process at a relatively small height (40 nm). This sort of measurements was acquired in a regime close to standard MFM technique. As temperature increases (7.53, 7.59, 7.66, and 7.72 K), we observe some instabilities resulting in abrupt changes in the phase shift. Images captured at temperatures from 7.79 to 7.85 K show that the magnetostatic interaction between the MFM tip and the vortex (in the lateral direction) approaches to the pinning interaction between the vortex and the defects, since visible shape of the vortex and its location are constantly changing. At the same temperature interval we observe the emergence of fragmentary formations reminiscent of a fish-scale motif (see inset in the image labelled 7.85 K). Further increase in temperature makes impossible to see individual pinned vortices, while the emerging fish-scale pattern expands until it entirely fills the scanned area (8.26 and 8.32 K). This pattern can be associated with nano-network of grains in the considered sample. It should be emphasized that successful visualizing the nano-network requires suppression of bulk pinning effects to allow the moving MFM probe to drag a single vortex during scanning. Since such dragged vortex can be considered as a nanoscale probe exploring the pinning potential of the sample, we refer to this regime as a scanning quantum-vortex microscopy. It is clear that the SQVM regime is effective in a limited temperature range close to $T^{\,}_c$. One fascinating aspect associated with the SQVM measurements is their unexpectedly high spatial resolution. Specifically, changes in phase manifest across a length scale of approximately $\sim 35\,$nm, what is significantly shorter than other lengths involved in the experiment (such as the effective magnetic penetration depth, size of the MFM probe, and scanning height).

\medskip

Let's examine the distinctive features of the phase-shift maps displaying a fish-scale pattern acquired in the SQVM regime at high temperatures in more detail. Figure~\ref{Fig04-comparison} show simultaneously recorded maps of the phase shift $\delta \varphi(x,y)$ (panel a) and the spatial variations of the amplitude of the oscillations $\delta A(x,y)$ (panel b). According to Eqs. (\ref{Eq:NC-AFM-Ampl}) and (\ref{Eq:NC-AFM-Phase}), both the amplitude of the mechanical oscillations at the resonant frequency and its phase could generally depend on the first-order derivative of the normal component of the force via the factor $F'_z$  and on local dissipation in the coupled system 'vortex-MFM probe' via the $Q$-factor. It is interesting to note that there is a partial correlation between the positions of the edges of the 'flakes' in the phase-shift map and the positions of the local minima of the amplitude (Fig.~\ref{Fig04-comparison}c). To determine which of these alternatives has a more significant impact, we show figure~\ref{Fig13-df} displaying the spatial variations of the frequency $\delta f(x,y)$ recorded with the regime with activated frequency-controlled feedback loop. We realize that the quality of this figure is not very high as compared to the phase-shift maps (see figures \ref{Fig04-evolution} and \ref{Fig04-comparison}a). Nevertheless, both panels in figure~\ref{Fig13-df} clearly demonstrate the presence of a fish-scale pattern even in this phase-locked loop mode. Since the frequency shift is proportional to the parameters of the force [see Eq. (\ref{Eq:NC-AFM-Freq})], we conclude that the fish-scale patterns visible in the SQVM mode is likely associated with spatial variations of the gradient of the $z$-component of the magnetostatic force. Such variations of $F'_z(x,y)$ arise as a dragged vortex moves through areas with strong and weak bulk pinning. In any case, successfully acquiring and comparing the maps recorded in the regimes with activated and deactivated frequency-controlled feedback would help us better understand the origin of the remarkable patterns.

The possible correspondence between the maps $\delta\varphi(x,y)$ and $\delta f(x,y)$ and the parameters of the local pinning allows us to interpret surprisingly high spatial resolution in the SQVM regime. Indeed, the most effective pinning relates to the case when the vortex core, whose electronic properties are similar to a normal metal, interacts with non-superconducting defects. According to our estimates, the radius of the vortex core at $T \gtrsim 8$\,K is about 36 nm. The grain boundaries are typically composed of disordered and partially oxidized Nb and thus they can serve as potential wells for vortex cores. This explains surprisingly high spatial resolution ($\gtrsim 35\,$nm) in the SQVM mode, compared with the superconducting coherence length at given temperature.

\medskip
\noindent \textsf{\textbf{50- and 100-nm thick Nb films.}} To confirm that the networks displayed in Figs.~\ref{Fig04-evolution}--\ref{Fig13-df} are not artefact of measurements, we extended our MFM/SQVM studies to thinner Nb films.  Essentially, the conclusions drawn from 240-nm-thick Nb films remain valid for 50- and 100-nm thick Nb films.

\begin{figure*}[ht!]
\centering
\includegraphics[width=150 mm]{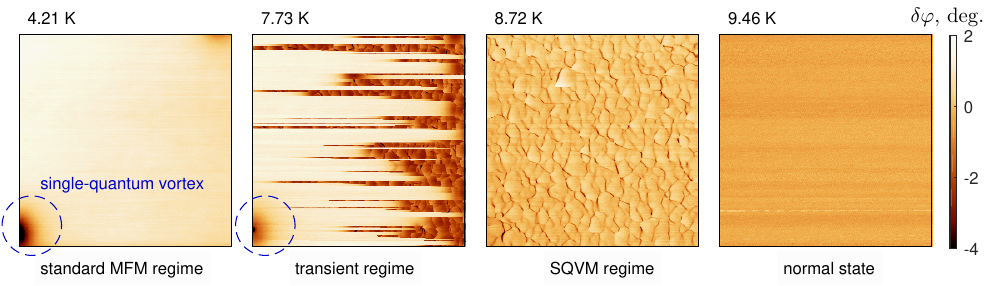}
\caption{Evolution of the phase-shift patterns $\delta\varphi(x,y)$ for the same area of 100-nm-thick Nb film with changing temperature (image size $5\times 5$ nm$^2$, scanning height 40 nm, regime of the constant frequency 48.178 kHz). }
\label{Fig-100nm-evolution}
\end{figure*}

\begin{figure}[ht!]
\centering
\includegraphics[width=95 mm]{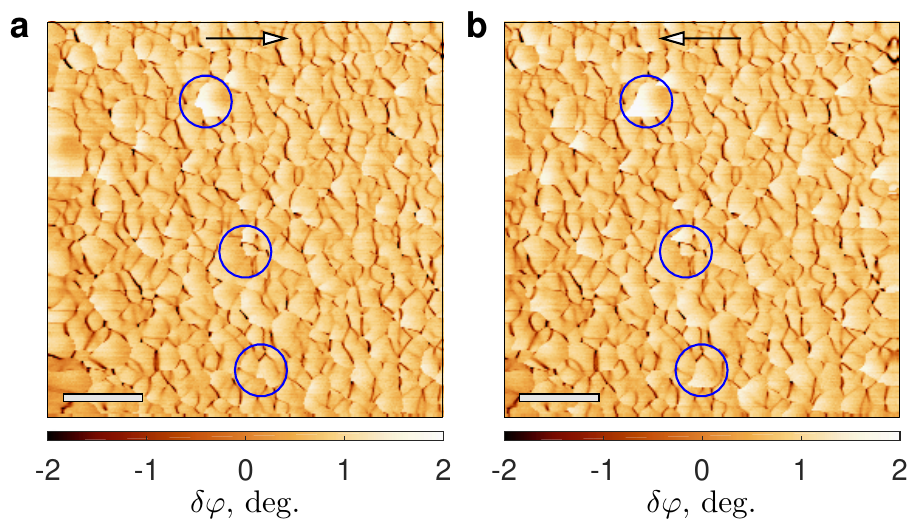}
\caption{\textbf{a, b} -- Comparison of the phase-shift maps $\delta\varphi(x,y)$ acquired for 100-nm thick Nb film in the forward and backward scanning directions (panels a and b, respectively) (image size $5\times 5$ nm$^2$, $T=8.64\,$K, scanning height 40\,nm, regime of the constant frequency 48.178 kHz). The direction of scanning are marked by arrows. Scale bars correspond to $1\,\mu$m. Although the majority of the fish-scale pattern remains consistent regardless of scanning direction, certain areas (indicated by red circles) exhibit hysteresis-like behavior.}
\label{Fig04-direction-scanning}
\end{figure}

\begin{figure*}[ht!]
\centering
\includegraphics[width=150 mm]{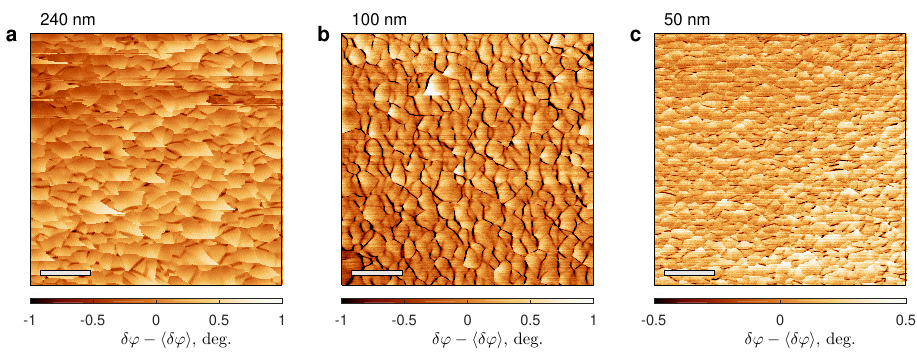}
\caption{\textbf{a, b} -- Comparison of the phase-shift maps $\delta\varphi(x,y)$ acquired for the Nb films of various thicknesses: 240 nm (panel a), 100 nm (panel b) and 50 nm (panel c).  Scale bars correspond to $1\,\mu$m (image size $5\times 5$ nm$^2$, $T=8.53\,$K (a), 8.72\,K (b) and 7.93\,K (c), scanning height 80\,nm, regime of the constant frequency 66.937 kHz (a), 48.178 kHz (b) and 62.804 kHz (c)).}
\label{Fig14}
\end{figure*}

Figure \ref{Fig-100nm-evolution} presents a sequence of phase-shift maps for a 100-nm-thick film at various temperatures. As expected, we can see different regimes: a conventional MFM regime displaying a single pinned vortex, a transient regime and and finally the SQVM regime revealing the underlying pinning landscape. The temperature interval within which the SQVM regime is expected broadens as the sample thickness decreases.\cite{Hovhannisyan-25}.

Notably, once the pinning landscape emerge, it remains largely unaffected by experimental parameters such as temperature, scanning speed and height, and direction of scanning. Two phase-shift maps, acquired sequentially line-by-line scanning in the forward and backward fast scanning directions, are compared in figure~\ref{Fig04-direction-scanning}. It is easy to see that these images closely resemble each other, suggesting that the fish-scale patterns are highly robust and largely unaffected by factors such as scanning direction, speed, or lift. Nevertheless, some localized differences -- spatial hysteresis and jumps -- can be detected (several instances highlighted by blue circles). Detailed studies of such instabilities depending on the scanning direction, as well as the development of theory considering different regimes of vortex pinning and depinning, are outside the scope of the current manuscript.

Figure \ref{Fig14} compares nano-networks presumably associated with pinning landscape in the Nb films of different thickness (50, 100 and 240 nm). Consistent with expectations, the average grain size enlarges as the film thickness increases.

\medskip
\noindent \textsf{\textbf{Pinning on defects vs. magnetic pinning.}} It is worth repeating that successful implementation of the SQVM regime requires the single-quantum vortex to first become unpinned from structural defects and then be guided by the MFM probe. It is clear that the attractive force between the vortex and the MFM probe can be enhanced by decreasing the scanning height, whereas the pinning potential due to structural defects can be lowered by increasing the sample temperature.

\begin{figure}[ht!]
\centering
\includegraphics[width=85 mm]{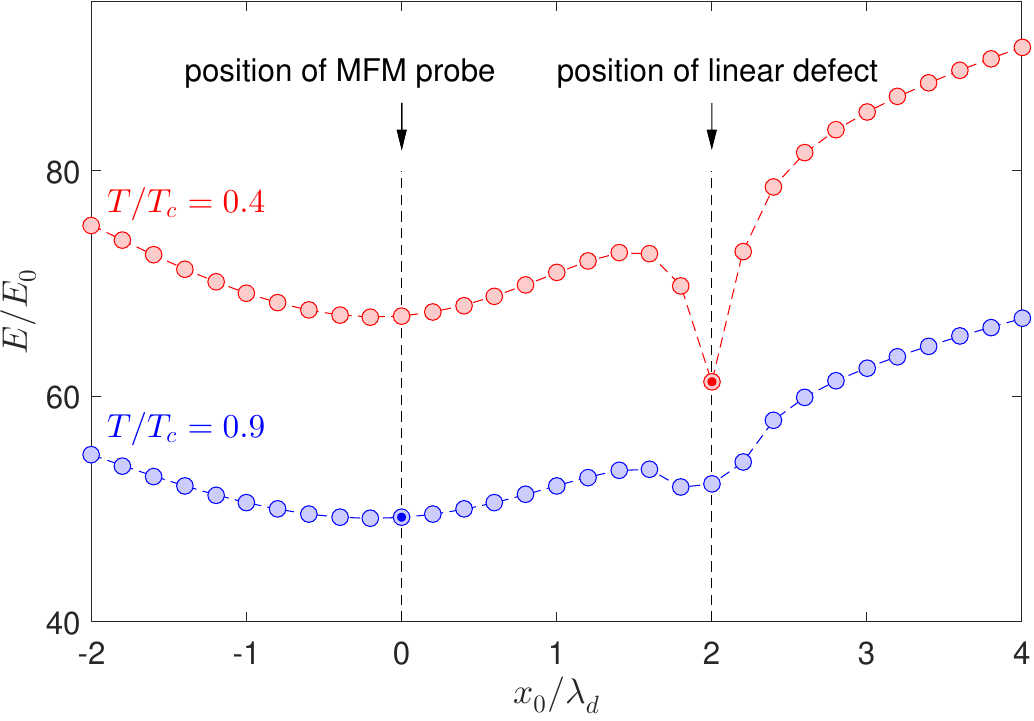}
\caption{Calculated dependence of the energy of the Pearl vortex $E$ in thin superconducting film as a function of the position of the vortex $x^{\,}_0$, where $E^{\,}_0 = \Phi^2_0/(64\pi^3 \lambda^{\,}_d(0))$. The position of the MFM probe and the linear defect, considered as an extended Josephson junction, are marked by dashed vertical lines.}
\label{Fig15}
\end{figure}

In order to confirm this expectations, we have performed a numerical modeling of the vortex pinning on a linear defect in thin superconducting film in the stationary London approximation. The modeling and calculation details are briefly described presented in Ref. \cite{Hovhannisyan-25}. In particular, it was shown that the dependence of the free energy $E$ of the Pearl vortex on its location $x^{\,}_0$ may generally exhibit two distinct minima (figure~\ref{Fig15}). One minimum describes the pinning of the vortex onto the structural defect, while the second minimum, located below the MFM probe, signifies the capture of the vortex by the probe's magnetic field. The relative depths of these minima vary as a function of temperature. Consequently, at lower temperatures, the structural defect exhibits stronger pinning compared to the magnetic interaction with the MFM probe, while at higher temperatures this relationship becomes inverted. Based on our simplified model, we could reasonably expect that the crossover temperature $T^*$, which separates two contrasting regimes of vortex pinning, should approach to $T^{\,}_c$ as the sample thickness and/or the scanning height increase. This finding aligns well with our experimental observations. We realize that such the London-based model neglects the influence of the normal-state vortex core in the energy balance, and therefore it cannot correctly describes all effects. Developing a more accurate model represents an independent challenge.


\noindent \textsf{\textbf{AFM mode vs. SQVM mode.}} In our recent publication\cite{Hovhannisyan-25} we have compared room-temperature AFM images with low-temperature SQVM images. Interestingly, small-scale  granular structure observed on the sample surface does not become apparent in the SQVM images  (see figures 1 and 2 in Ref.~\cite{Hovhannisyan-25}). As previously mentioned, the SQVM measurements should be sensitive primarily to the presence of pinning-effective defects within the material, rather than to surface irregularities. It points to that the SQVM approach can provide valuable information about hidden defects in thin superconducting films, not accessible by other experimental techniques (AFM, MFM, STM/STS etc).


\section*{Conclusion}

We have developed a simple experimental technique -- scanning quantum-vortex microscopy -- for deeper investigations of the vortex pinning nano-network in granular superconducting films with high spatial resolution. The method is based on creating and dragging a single-quantum vortex by the probe of a magnetic force microscope. The interaction of the dragged vortex with structural defects within thin-film sample evidently induces a position-dependent pinning force, altering the net force between the vortex and the MFM probe. Such changes can be easily detected by examining additional shifts in the amplitude ($\delta A$), resonant frequency ($\delta f$), or phase ($\delta \varphi$) of the driven oscillations of the MFM cantilever, operating in a non-contact mode. Since the vortex line fully penetrates the superconducting film, the dragged vortex can be regarded as a nano-scale probe sensitive to both surface and bulk components of the local pinning potential. Spatial variations of the frequency or phase of the MFM probe oscillations can be presented in a form of two-dimensional maps: $\delta f(x,y)$ and $\delta \varphi(x,y)$. We believe that such maps depict the projections of the subsurface grain boundaries and other types of extended defects inside the superconducting film, which are effective for vortex pinning, onto the scanning plane. This technique has been successfully applied to conventional magnetron-sputtered niobium films, which find extensive application in contemporary superconducting electronics. We have demonstrated the SQVM approach facilitates the visualization of hidden defects in thin-film Nb samples with nanometer resolution over a large field of view. This opens exciting possibilities for characterization and non-destructive studies of subsurface defects inside thin superconducting films (NbN, TiN and others), nanodevices and hybrid structures. It is important to note that an appropriate theoretical model correctly describing all relevant processes associated with viscous motion of the Pearl vortex in spatially-inhomogeneous pinning potential under the influence of an oscillating MFM probe has not been fully developed yet. This problem, which has direct practical implications, may be of interest to theorists specializing in the field of superconductivity.


\section*{Acknowledgements}

All MFM/SQVM experiments and their analysis were carried out with the support of the Russian Science Foundation (project No. 23-72-30004). The samples were fabricated with the support of the Ministry of Science and Higher Education of the Russian Federation (project No. 075-15-2025-010).

\section*{Data availability}

\noindent
The data that support the findings of this study are available from the corresponding authors upon request.


%
%
%

\section*{Author contribution}

Vasily Stolyarov proposed the concept for the experiment and supervised the experiments. Andrey Shishkin, Olga Skryabina and Vasily Stolyarov prepared the experimental samples; Razmik Hovhannisyan, Sergey Grebenchuk,  Semyon Larionov, and Vasily Stolyarov carried out low-temperature MFM/SQVM measurements; Razmik Hovhannisyan, Simon Larionov, and Alexey Yu. Aladyshkin performed numerical analysis of the MFM and SQVM data; Dimitry Roditchev, Alexey Aladyshkin and Vasily Stolyarov provided the explanation of the observed effects; Alexander Mel'nikov and Alexey Samokhvalov did numerical modeling; Alexey Aladyshkin and Vasily Stolyarov drafted the manuscript, incorporating critical input from all co-authors.

\section*{Competing Interests}
The authors declare no competing financial or non-financial interests.


\begin{references}
\bibitem{Comment}
Any superconducting sample formally frozen in zero applied magnetic field can have vortices due to Earth's residual magnetic field (about 0.5 Oe). Our MFM measurements shows that $7\times 7\,\mu$m$^2$ areas of zero-field-cooled sample typically contain one or two vortices (see figure \ref{Fig-240-df-dphi}). This observed value aligns well with straightforward calculations predicting the expected number of single-quantum vortices: $N=\Phi/\Phi^{\,}_0\simeq 1.2$, where $\Phi = H\cdot S = 2.5\cdot 10^{-7}$ Oe$\cdot$cm$^2$ is the flux of the Earth's magnetic field through the sample's surface, $\Phi^{\,}_0 = 2\cdot 10^{-7}$ Oe$\cdot$cm$^2$ is the magnetic flux quantum.
\end{references}
\end{document}